\documentclass{INTERSPEECH2023}

\usepackage{epstopdf}
\usepackage{subfig}
\usepackage{graphicx}
\interspeechcameraready

\title{Non-uniform Speaker Disentanglement For \\ Depression Detection From Raw Speech Signals}
\name{Jinhan Wang$^1$, Vijay Ravi$^1$, Abeer Alwan$^1$ \thanks{This work was funded by NIH award number R01MH122569.}}
\address{
  $^1$Dept . of Electrical and Computer Engineering, University of California, Los Angeles, USA}
\email{(wang7875@g, vijaysumaravi@g, alwan@ee).ucla.edu}

\begin{document}

\maketitle
 
\begin{abstract}
\noindent While speech-based depression detection methods that use speaker-identity features, such as speaker embeddings, are popular, they often compromise patient privacy. To address this issue, we propose a speaker disentanglement method that utilizes a non-uniform mechanism of adversarial SID loss maximization. This is achieved by varying the adversarial weight between different layers of a model during training. We find that a greater adversarial weight for the initial layers leads to performance improvement. Our approach using the ECAPA-TDNN model achieves an F1-score of 0.7349 (a 3.7\% improvement over audio-only SOTA) on the DAIC-WoZ dataset, while simultaneously reducing the speaker-identification accuracy by 50\%. Our findings suggest that identifying depression through speech signals can be accomplished without placing undue reliance on a speaker's identity, paving the way for privacy-preserving approaches of depression detection. 
\end{abstract}


\section{Introduction}
\label{sec:intro}
Speech signals have emerged as significant biomarkers of one's emotional and mental state~\cite{cummins2015review, nilsonne1988speech, andreasen1976linguistic, ravi2019voice}. Several previous studies have successfully demonstrated the potential of using speech in developing automatic objective screening systems for mental health disorders, including serious illnesses such as Major Depressive Disorder (MDD)~\cite{alghowinem2013detecting,ringeval2019avec,low2020automated}. Various features and model architectures have been proposed in the past for the purpose of MDD diagnosis~\cite{rejaibi2019mfcc, shen2022automatic,chlasta2019automated}, each having its own distinct set of advantages and limitations. These include spectral~\cite{sanchez2011using,dubagunta2019learning}, prosodic~\cite{yang2012detecting}, voice quality~\cite{afshan2018effectiveness} and articulatory~\cite{seneviratne2022multimodal} features as well sophisticated modeling techniques such as data augmentation~\cite{yang2020feature}, model ensembles~\cite{vazquez2020automatic}, transfer-learning~\cite{harati2021speech} and self-supervised pre-training~\cite{wang22z_interspeech}. 

Similarly, speaker-identity-related features have been used in depression detection, with previous studies focusing on i-vectors~\cite{di2021using}, x-vectors~\cite{egas2022automatic, ravi2022fraug}, or speaker embeddings~\cite{dumpala2022detecting}. Although these models result in good performance, the use of speaker-identity-related features raises privacy-preservation concerns. In the healthcare system, where there is often a stigma surrounding mental health, it is important to develop models that are less reliant on an individual's identity~\cite{lustgarten2020digital}.

Although the field of privacy-preserving depression detection is relatively new, a few studies have attempted to endeavor in this direction. Among them, Federated learning~\cite{suhasprivacy} and sine-wave speech~\cite{dumpala21_smm} are notable examples of such methods. Although these methods are promising, their application to low-resource depression detection from speech signals is still in its early stages, and results in significant performance loss~\cite{suhasprivacy}.

More recently,~\cite{li2020speaker,gat2022speaker,yin2020speaker} proposed to remove speaker-related information from speech signals using adversarial learning for speech emotion recognition. We refer to this approach as uniform speaker disentanglement (USD) where the whole model is trained with the same adversarial loss. Despite the promising results of USD in detecting depression, as reported in~\cite{ravi22_interspeech}, the model has certain limitations that can impede its performance. One such limitation is the lack of consideration for the interactions between different layers of the model and the relationship between the tasks being performed and the intermediate representations. For example, recent research has shown that different layers of a model capture information differently~\cite{chen2022wavlm}. It is, therefore, possible that some layers capture more depression information and less speaker information or vice versa, and applying speaker disentanglement to all the layers uniformly may result in sub-optimal performance.

In this paper, we hypothesize that speaker-related information encoded by different layers of a model is idiosyncratic, both in terms of quantity and quality, where some layers may encode more or fewer speaker characteristics than other layers, some of which may not be relevant for depression detection. Assigning a higher penalty to such layers during adversarial training can improve overall model performance. Hence, we propose a novel non-uniform speaker disentanglement method (NUSD) that regulates the proportion of speaker disentanglement applied to different model layers and show that NUSD outperforms USD. 

We introduce a new model-input combination by training the ECAPA-TDNN model~\cite{desplanques2020ecapa} with raw-audio speech signals as input. NUSD is implemented by adjusting the weighting of the adversarial loss between the two components of the model: the feature extraction (FE) and the feature processing (FP) sections. This method achieves audio-only state-of-the-art (AO-SOTA) performance on depression detection using DAIC-WoZ dataset~\cite{valstar2016avec} while simultaneously lowering speaker identification (SID) accuracy. We analyze the behavior of the model layers using a class separability framework, finding that a higher adversarial weight to the FE layers more effectively suppresses speaker information than USD, leading to a better encoding of depression information and performance improvement. To the best of our knowledge, our study is the first to suggest the use of a layer-behavior-based manipulation of loss, in that, we 1) propose differential weighting of the adversarial loss, and 2) utilize the functionality of the FE and FP layers to decide on weight distribution. 

The paper is structured as follows: Section~\ref{sec:nusd} presents the proposed NUSD method. Datasets, models, and experiments are described in Section~\ref{sec:exp_details}. Results are discussed in Section~\ref{sec:results}, and Section~\ref{sec:conclusion} concludes the paper and outlines future directions.


\section{Speaker Disentanglement}
\label{sec:nusd}
Uniform speaker disentanglement (\textbf{USD})~\cite{ravi22_interspeech} minimizes the prediction loss for the primary task and maximizes the loss of the auxiliary task. In the context of this paper, the primary task is depression detection, and the auxiliary task is SID. Consequently, the USD loss function is - 
\begin{equation}
\label{eq:USD_loss}
    L_{USD} = L_{MDD} - \lambda ( L_{SPK}) 
\end{equation}

where $L_{MDD}$ is the depression-detection loss and $\lambda$ controls how much of the SID loss, $L_{SPK}$ contributes to the total loss, $L_{USD}$. Conventionally, $L_{MDD}$ is Binary Cross Entropy loss, and $L_{SPK}$ is multi-class Cross-Entropy loss. 

A higher value of $\lambda$ indicates a greater adversarial cost during training. This in turn scales the speaker-loss gradient of all the layers uniformly, by the same factor $\lambda$. Let the trainable parameters of a model be denoted as $\theta_{ALL}$, then, the gradient of $L_{SPK}$ in USD can be expressed as - 

\begin{equation}
\label{eq:USD_gradients}
    \frac{\partial L_{SPK}(_{USD})}{\partial \theta_{ALL}} = \frac{\partial( \lambda L_{SPK})}{\partial \theta_{ALL}}
\end{equation}

During the optimizer's update step, the model's parameters are modified as follows:
\begin{equation}
\label{eq:USD_optimiization}
    \theta_{ALL} = \theta_{ALL} + \alpha ( \frac{\partial L_{SPK}}{\partial \theta_{ALL}} -  \frac{\partial L_{MDD}}{\partial \theta_{ALL}} )
\end{equation}

where $\alpha$ is the learning rate. The negative term (positive sign) for the speaker gradient in Eq.~\ref{eq:USD_optimiization} ensures that the model maximizes $L_{SPK}$ while simultaneously optimizing $L_{MDD}$ thereby partially disentangling speaker identity and depression status. Although USD has shown promising results in speaker disentanglement for depression detection~\cite{ravi22_interspeech}, it can be further improved by providing better control over the proportion of adversarial disentanglement applied to different model layers. 

To address this limitation, we propose a non-uniform speaker disentanglement (\textbf{NUSD}) approach. The idea is as follows - the loss gradients of the auxiliary task can be split into multiple components based on model layers and unlike USD, loss maximization can be applied differently to each component thereby allowing for varying levels of disentanglement to be applied to different layers. 

As a preliminary study, we split the gradients into two components: the feature extraction component (FE) composed of the initial layers, and the feature processing component (FP) made up of the final layers as detailed in Section~\ref{subsubsec:nusd_components}. The trainable parameters of these components can be represented as $\theta_{FE}$ and $\theta_{FP}$ for the FE \& the FP layers, respectively. The speaker-loss gradients of FE and FP are non-uniformly scaled using different factors $\lambda_1$ and $\lambda_2$, respectively. Therefore, for NUSD, the gradient of $L_{SPK}$ can be written as -  
\begin{equation}
\label{eq:NUSD_gradients}
    \frac{\partial L_{SPK}(_{NUSD})}{\partial \theta_{ALL}} = [  \frac{\partial( \lambda_1 L_{SPK})}{\partial \theta_{FE}},  \frac{\partial( \lambda_2 L_{SPK})}{\partial \theta_{FP}}]
\end{equation}   

Comparing Eq.~\ref{eq:NUSD_gradients} with Eq.~\ref{eq:USD_gradients}, it can be observed that NUSD helps us regulate adversarial disentanglement of different layers of the model differently by changing the ratio of $\lambda_1$ to $\lambda_2$ (denoted as $\beta$ in later sections). For example, if $\beta < 1$ i.e., $\lambda_2 > \lambda_1$, then the FP layers are penalized more than the FE layers during adversarial training and vice-versa. Conversely, if $\beta=1$, then NUSD is equivalent to USD.

\section{Experimental Details}
\label{sec:exp_details}
This section outlines experimental details, including the dataset, preprocessing steps, and the model architecture. Two models, ECAPA-TDNN~\cite{desplanques2020ecapa} and DepAudioNet~\cite{ma2016depaudionet} were trained on a publicly available dataset to showcase our approach's effectiveness. ECAPA-TDNN is a SOTA model in SID~\cite{desplanques2020ecapa} and emotion recognition~\cite{speechbrain, Morais_ser_ecapa}, while DepAudioNet is a common depression literature baseline~\cite{ma2016depaudionet}.

\subsection{Dataset and Input Features}
\subsubsection{Database}
The DAIC-WoZ database~\cite{valstar2016avec} is a collection of audio-visual interviews in English featuring 189 participants, both male, and female, who underwent psychological distress evaluations. The dataset contains 107 speakers used for training and 35 speakers used for evaluation, consistent with the database description. Audio data from only the patients were extracted using the provided time labels. The validation set was utilized to report results, in line with previous literature. 

\subsubsection{Data Pre-processing and Input features}
Models were trained using raw-audio features as input, with pre-processing steps implemented to address data imbalance~\cite{ma2016depaudionet,bailey2021gender,ravi22_interspeech,wang22z_interspeech}. Prior to training, the training data were pre-processed with random cropping and sampling, where each utterance was randomly cropped to the length of the shortest utterance and segmented into multiple 3.84s segments (equivalent to 61440 raw-audio samples). To generate a balanced training subset, an equal number of depression and non-depression segments were randomly sampled without replacement. In each experiment, five models were trained using a randomly generated training subset, with the final prediction averaged across the five models. The raw-audio samples were normalized using mean-variance normalization~\cite{bailey2021gender}.

\begin{table}[hb]
\centering
\caption{Architecture details of the ECAPA-TDNN Model. [InC,OutC,K,S,P,D] are in-channels, out-channels, kernel, stride, padding and dilation, respectively.}
\label{tab:ecapa-architecture}
\resizebox{0.8\linewidth}{!}{%
\begin{tabular}{lr}
\hline 
\textbf{Layer Name} & \textbf{InC,OutC,K,S,P,D} \\
\hline 
Input Layer & [1,128,1024,512,0,1] \\
SE-Res2-1 & [128,128,3,1,2,2] \\
SE-Res2-2 & [128,128,3,1,3,3] \\
SE-Res2-3 & [128,128,3,1,4,4] \\
\hline
Feature aggregation & - \\
Concat-Conv & [384,384,1,1,0,1] \\
AttentiveStatsPool & [384,768,-,-,-,-] \\
Embedding Layer & [768,128,-,-,-,-] \\
Speaker Prediction Layer & [128,107,-,-,-,-] \\
Depression Prediction Layer & [128,1,-,-,-,-] \\
\hline 
\end{tabular}}
\end{table}

\subsection{Models}
\label{subsec:ECAPA_exp_details}

\subsubsection{ECAPA-TDNN}

In contrast to previous studies~\cite{wang22q_interspeech} that use spectrograms or MFCCs as inputs, the proposed ECAPA-TDNN model is trained using raw-audio signals. The model architecture was modified (see Table~\ref{tab:ecapa-architecture}) to accommodate raw-audio speech signals as input and avoid overfitting the model to a small training dataset.  Specifically, the kernel and stride of the input convolution layer, the number of channels in the intermediate layers, and the dimensions of the prediction layers were modified. 

\subsubsection{DepAudioNet}
This model employs a CNN-LSTM architecture ~\cite{ma2016depaudionet} with implementation based on~\cite{bailey2021gender}.
Two 1-D Convolution layers followed by two unidirectional LSTM layers were used. Lastly, the MDD and speaker prediction layers were fully connected layers with output dimensions for speaker labels being 107.

\subsection{Experiments}
\subsubsection{USD and NUSD}
\label{subsubsec:nusd_components}
Both models share the same adversarial weight $\lambda$ across all layers in the USD experiments. In contrast, in the NUSD experiments, the FE layers are weighted with $\lambda_1$ and the FP layers with $\lambda_2$. We consider the input layer and three SE-Res2 blocks of the ECAPA-TDNN model as FE, while the feature aggregation layer, concat-Conv layer, attention layer, fully connected embedding layer, and prediction layers are FP. Similarly, for DepAudioNet, the first 2 convolutional layers are FE with the two LSTM layers along with the prediction layers as FP. $\beta$ and $\lambda$ values are empirically chosen\footnote{Code repository: https://github.com/kingformatty/NUSD}.

\subsubsection{Speaker Identification Experiments}
To investigate how speaker disentanglement affects speaker identity, we conduct a SID experiment by training a support vector classifier (SVC) using speaker embeddings (the embedding layer output from the ECAPA-TDNN model or the hidden representation of the last LSTM layer from the DepAudioNet model). During SVC training, embeddings are obtained from the baseline model without speaker disentanglement, while the evaluation embeddings are taken from the model with or without speaker disentanglement. Note that the SID branch of the model is discarded when extracting speaker embeddings.

\subsubsection{Layer-wise GDV Analysis}
Because the proposed method regulates the magnitude of adversarial disentanglement applied to different components of the models, we investigate the layer-wise behavior of the models with and without NUSD. This is accomplished with Generalized Discrimination Value (GDV)~\cite{schilling2021quantifying} analysis. Previously, GDV has been proposed as a metric to evaluate the separability of specific representations with respect to various classes and data labels. In this paper, we employ GDV to measure the speaker and MDD-separability of individual layers' outputs for the models in consideration. The prediction layers are excluded in this analysis and GDV values are sign-flipped, such that a higher GDV stands for better separability. 

\section{Results and Discussion}
\label{sec:results}
Results are shown in Table~\ref{tab:main_results_summary}, where the best results from the literature are in the first part, and the baselines and proposed method, are in the second part. Methods are compared using speaker-level F1-scores for the depressed (F1-D), the non-depressed (F1-ND) classes, and their non-weighted (macro) average (F1-AVG). To measure the degree of speaker disentanglement, the accuracy of SID was reported if applicable. 

\begin{table*}[htbp]
\centering
\caption{Depression detection performance for various models and AO-SOTA baselines based on F1-AVG, F1(ND), F1(D), and Speaker ID accuracy using the DAIC-WoZ dataset. SOTA baseline results are either reproduced values or reported from the corresponding study. The symbol `$-$' indicates that those values were not reported in the corresponding study. The symbols `~$\uparrow$' and `~$\downarrow$' indicate a higher or lower value is better, respectively.  Best results are highlighted in bold.}
\label{tab:main_results_summary}
\resizebox{0.9\linewidth}{!}{%
\begin{tabular}{llccrrrr} 
\hline 
\textbf{Model Architecture} & \textbf{Input Feature} & \begin{tabular}[c]{@{}c@{}}\textbf{Disentanglement}\\\textbf{Method}\end{tabular} & \begin{tabular}[c]{@{}c@{}}\textbf{Model}\\\textbf{Parameters}~\end{tabular} & \textbf{F1-AVG~$\uparrow$} & \textbf{F1(ND)~$\uparrow$} & \textbf{F1(D)~$\uparrow$} & \begin{tabular}[c]{@{}c@{}}\textbf{SID}\\\textbf{Accuracy}\end{tabular}~$\downarrow$ \\ 
\hline 
DepAudioNet~\cite{ma2016depaudionet} & Mel-Spectrogram & None & 280k & 0.6081 & 0.6977 & 0.5185 & - \\

FVTC-CNN~\cite{huang2020exploiting} & Formants & None & - & 0.6400 & 0.4600 & 0.8200 & - \\
Speech SimCLR~\cite{jiang2020speech} & Mel-Spectrogram & None & - & 0.6578 & 0.7556 & 0.5600 & - \\
CPC~\cite{oord2018representation} & Mel-Spectrogram & None & - & 0.6762 & 0.7317 & 0.6207 & - \\
CNN-LSTM~\cite{dumpala2022detecting} & Spk. Embd. + OpenSmile & None & - & 0.6850 & \textbf{0.8600} & 0.5100 & - \\
SpeechFormer~\cite{chen22_interspeech} & Wav2Vec & None & 33M & 0.6940 & - & - & - \\ 
Vowel-based~\cite{feng2022toward} & Mel-Spectrogram & None & - & 0.7000 & 0.8400 & 0.5600 & - \\\hline
DepAudioNet~\cite{bailey2021gender} ($D1$) & Raw-Audio & None & 445k & 0.6259 & 0.7755 & 0.4762 & 10.04\% \\
DepAudioNet~\cite{ravi22_interspeech} ($D2$) & Raw-Audio & USD & 459k & 0.6830 & 0.7826 & 0.5833 & 8.91\% \\
DepAudioNet ($D3$) & Raw-Audio & NUSD & 459k & 0.7086 & 0.8085 & 0.6087 & 8.05\% \\
ECAPA-TDNN ($E1$) & Raw-Audio & None & 595k & 0.6329 & 0.7273 & 0.5385 & 42.33\% \\

ECAPA-TDNN ($E2$) & Raw-Audio & USD & 609k & 0.7086 & 0.8085 & 0.6087 & 9.38\% \\

ECAPA-TDNN ($E3$) & Raw-Audio & NUSD & 609k & \textbf{0.7349} & 0.8333 & \textbf{0.6364} & \textbf{4.68\%} \\ 
\hline
$\Delta$ ($E3$ vs $E2$) in \% & - & - & - & 3.70 & 2.80 & 4.55 & -50.11 \\
\hline
\end{tabular}
}

\end{table*}

\subsection{Baseline Experiments}
The DepAudioNet model ($D1$), trained on raw-audio, achieves an F1-Score of 0.6259, whereas the proposed ECAPA-TDNN model ($E1$), also trained on raw audio signals and without speaker disentanglement, achieves an F1-Score of 0.632, demonstrating a 1.12\% improvement. Some previous studies have achieved better results than $D1$ and $E1$, for example the Vowel-based study~\cite{feng2022toward} (0.7) and the SpeechFormer~\cite{chen22_interspeech} (0.694). However, these studies have certain limitations. The Vowel-based study requires a trained vowel classification model, while SpeechFormer is a large model with 33M parameters. In contrast, the proposed raw audio-based ECAPA-TDNN model has only 609k parameters and does not need any auxiliary classifiers nor expensive self-supervised models making it simpler and more efficient. 


\subsection{Speaker Disentanglement}
When USD is applied to the DepAudioNet Model ($D2$), its performance improves by 9.12\% from 0.6259 to 0.6830 ($\lambda = 3\mathrm{e}{-4}$). Furthermore, when the ECAPA-TDNN model is trained using USD ($E2$), it achieves an impressive F1-Score of 0.7086 ($\lambda = 3\mathrm{e}{-3}$), outperforming $E1$ by 11.96\%. Along with a significant increase in MDD classification performance, there is a decrease in SID accuracy of 11.2\% (from 10.04\% to 8.91\%) and 77.8\% (from 42.33\% to 9.38\%) for $D2$ and $E2$, respectively. 

Next, we apply NUSD to the DepAudioNet and ECAPA-TDNN models and label the resulting best-performing models as $D3$ and $E3$, respectively. Model $D3$ achieves an F1 score of 0.7086 ($\lambda_1 = 2\mathrm{e}{-3}$, $\lambda_2 = 4\mathrm{e}{-4}$), an increase of 13.21\% over $D1$ and 3.75\% over $D2$, while only marginally reducing the SID accuracy to 8.05\%. The overall best-performing model is $E3$, which achieves an F1 score of 0.7349 ($\lambda_1 = 4\mathrm{e}{-5}$, $\lambda_2 = 8\mathrm{e}{-6}$), outperforming other AO-SOTA models in the literature and surpassing the corresponding baseline models $E1$ and $E2$ by 16.12\% and 3.7\%, respectively while simultaneously reducing the SID accuracy to 4.68\%. These results imply that applying NUSD can be an effective way to enhance depression classification performance while reducing SID performance.

The proposed $E3$ model surpasses AO-SOTA performance in depression detection without requiring additional training data, sophisticated pre-trained models, or complicated handcrafted features. Compared to some previously published AO-SOTA results, the method achieves an improvement of 4.98\% vs. vowel-based~\cite{feng2022toward} and 5.89\% vs. SpeechFormer~\cite{chen22_interspeech}. Although~\cite{dumpala2022detecting} has a better F1-ND when combining speaker embeddings with OpenSmile~\cite{eyben2010opensmile} features, our method achieves a better overall F1-AVG. Moreover,~\cite{dumpala2022detecting} uses a segment-level evaluation procedure, in contrast to using speaker-level as in this paper, and reports a lower F1-D of 0.43 and F1-ND of 0.82 when only speaker embeddings are used without feature-fusion.


In pilot experiments, we found that raw audio outperformed Mel-spectrograms, and ComparE16~\cite{eyben2010opensmile}. Hence, we used the strongest baseline as the goal of the work is to provide a framework that can improve performance regardless of the features chosen and not to compare the performance based on features. 
\vspace{-1mm}
\subsection{Ablation Experiments}

\begin{figure}[htbp]
    \centering
    \includegraphics[width=0.9\linewidth]{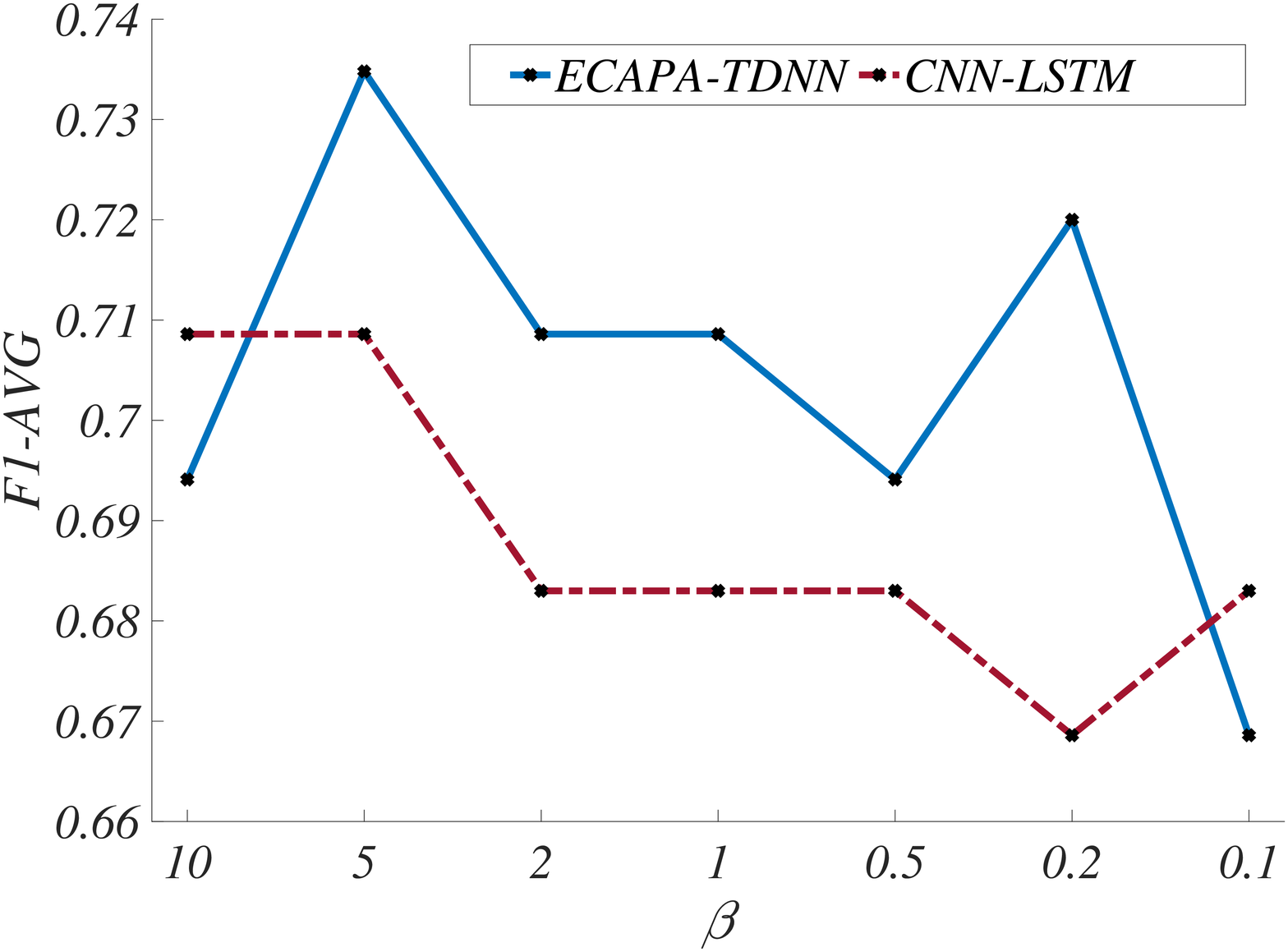}
    \caption{\label{fig:beta_vs_f1} A plot of F1-AVG versus NUSD $\beta$ values for the ECAPA-TDNN and the DepAudioNet CNN-LSTM model. Best viewed in color.}
\end{figure}

In order to study the impact of the hyperparameter $\beta$ on model performance, we conducted a series of experiments using different values of $\beta$ ranging from 10 to 0.1. The F1-AVG scores were plotted as a function of $\beta$ for both models (Figure~\ref{fig:beta_vs_f1}). Our analysis revealed two key observations: Firstly, for both models, NUSD ($\beta = 5$) consistently outperformed USD ($\beta = 1$), indicating that a non-uniform manner of adversarial training can be beneficial for performance.  

Secondly, we observed a trend in both DepAudioNet and ECAPA-TDNN models wherein higher values of $\beta$ produced better results up to $\beta = 5$. This finding suggests that assigning a higher penalty to the initial layers than to the final layers during adversarial training can improve model performance. One possibility is that assigning a higher weight to penalize FE layers in NUSD leads to more effective suppression of speaker-specific feature extraction that may not be too relevant to the primary task of depression detection, compared to assigning the same weight to both FE and FP layers as in USD. Although this is a consequential outcome and holds true for depression detection using the DAIC-WoZ dataset, further investigation is required to verify that the framework generalizes to other domains.

\subsection{Layer-wise GDV Analysis}
\begin{figure}[htbp]
    \hspace*{-0.65cm}
    \centering
    \includegraphics[width=1.2\linewidth, height=4.6cm]{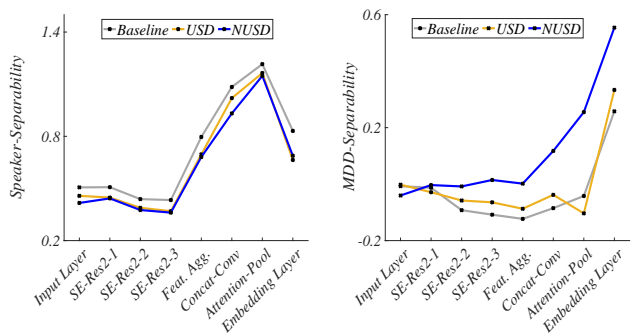}
    \caption{\label{fig:gdv-ecapa} A plot of the layer-wise speaker and  MDD separability GDV scores of the ECAPA-TDNN model. Three models are analyzed - baseline without speaker disentanglement ($E1$), USD ($E2$), and NUSD ($E3$). X-axis represents the layers of the ECAPA-TDNN model. Best viewed in color.}
\end{figure}

MDD and speaker separability of individual layers of the $E1$, $E2$, and $E3$ were analyzed using the GDV scores (Figure~\ref{fig:gdv-ecapa}).  Overall, these plots offer valuable insights into the behavior of the model and shed light on how the proposed method affects the separation of depression and speaker features within the model. Notably, for speaker-separability, NUSD had the lowest value among the three methods in all layers except in the embedding layer (0.664 for USD vs. 0.688 for NUSD) showing that NUSD was better at speaker disentanglement than USD in the FE layers and comparable to USD in the FP layers.  

For depression separability, we observe that NUSD has a significantly better separability profile throughout the model compared to USD and the baseline counterparts. These findings support our hypothesis that speaker information encoded by different layers of the model is distinctive and non-uniform speaker disentanglement, which exploits these characteristics of model behavior, leads to better depression detection.

\section{Conclusion}
\label{sec:conclusion}
The proposed privacy-preserving approach of speech-based depression detection shows promising results by utilizing a non-uniform mechanism of adversarial SID loss maximization. The approach achieves an F1-Score of 0.7349 on the publicly available DAIC-WoZ dataset without any data augmentation, pre-training, or handcrafted features, outperforming other AO-SOTA methods while simultaneously reducing SID accuracy. These findings suggest that our approach leads to better model performance with improved speaker disentanglement. Future work will focus on analyzing the effects of the number of speakers in the database, exploring a more fine-grained, data-driven variant of NUSD, and extending the approach to other domains.


\bibliographystyle{IEEEtran}
\bibliography{mybib}

\begin{thebibliography}{10}
\providecommand{\url}[1]{#1}
\csname url@samestyle\endcsname
\providecommand{\newblock}{\relax}
\providecommand{\bibinfo}[2]{#2}
\providecommand{\BIBentrySTDinterwordspacing}{\spaceskip=0pt\relax}
\providecommand{\BIBentryALTinterwordstretchfactor}{4}
\providecommand{\BIBentryALTinterwordspacing}{\spaceskip=\fontdimen2\font plus
\BIBentryALTinterwordstretchfactor\fontdimen3\font minus
  \fontdimen4\font\relax}
\providecommand{\BIBforeignlanguage}[2]{{%
\expandafter\ifx\csname l@#1\endcsname\relax
\typeout{** WARNING: IEEEtran.bst: No hyphenation pattern has been}%
\typeout{** loaded for the language `#1'. Using the pattern for}%
\typeout{** the default language instead.}%
\else
\language=\csname l@#1\endcsname
\fi
#2}}
\providecommand{\BIBdecl}{\relax}
\BIBdecl

\bibitem{cummins2015review}
N.~Cummins \emph{et~al.}, ``A review of depression and suicide risk assessment
  using speech analysis,'' \emph{Speech Communication}, vol.~71, pp. 10--49,
  2015.

\bibitem{nilsonne1988speech}
A.~Nilsonne, ``Speech characteristics as indicators of depressive illness,''
  \emph{Acta Psychiatrica Scandinavica}, vol.~77, no.~3, pp. 253--263, 1988.

\bibitem{andreasen1976linguistic}
N.~J. Andreasen \emph{et~al.}, ``Linguistic analysis of speech in affective
  disorders,'' \emph{Archives of General Psychiatry}, vol.~33, no.~11, pp.
  1361--1367, 1976.

\bibitem{ravi2019voice}
V.~Ravi \emph{et~al.}, ``Voice quality and between-frame entropy for sleepiness
  estimation,'' \emph{Interspeech}, 2019.

\bibitem{alghowinem2013detecting}
S.~Alghowinem \emph{et~al.}, ``Detecting depression: a comparison between
  spontaneous and read speech,'' in \emph{ICASSP}.\hskip 1em plus 0.5em minus
  0.4em\relax IEEE, 2013, pp. 7547--7551.

\bibitem{ringeval2019avec}
F.~Ringeval \emph{et~al.}, ``Avec 2019 workshop and challenge: state-of-mind,
  detecting depression with ai, and cross-cultural affect recognition,'' in
  \emph{Proceedings of the 9th AVEC}, 2019.

\bibitem{low2020automated}
D.~M. Low \emph{et~al.}, ``Automated assessment of psychiatric disorders using
  speech: A systematic review,'' \emph{Laryngoscope Investigative
  Otolaryngology}, vol.~5, no.~1, pp. 96--116, 2020.

\bibitem{rejaibi2019mfcc}
E.~Rejaibi \emph{et~al.}, ``Mfcc-based recurrent neural network for automatic
  clinical depression recognition and assessment from speech,''
  \emph{Biomedical Signal Processing and Control}, vol.~71, p. 103107, 2022.

\bibitem{shen2022automatic}
Y.~Shen \emph{et~al.}, ``Automatic depression detection: An emotional
  audio-textual corpus and a gru/bilstm-based model,'' in \emph{ICASSP}.\hskip
  1em plus 0.5em minus 0.4em\relax IEEE, 2022, pp. 6247--6251.

\bibitem{chlasta2019automated}
K.~Chlasta \emph{et~al.}, ``Automated speech-based screening of depression
  using deep convolutional neural networks,'' \emph{Procedia Computer Science},
  vol. 164, pp. 618--628, 2019.

\bibitem{sanchez2011using}
M.~H. Sanchez \emph{et~al.}, ``Using prosodic and spectral features in
  detecting depression in elderly males,'' in \emph{Interspeech}, 2011, pp.
  3001--3004.

\bibitem{dubagunta2019learning}
S.~P. Dubagunta \emph{et~al.}, ``Learning voice source related information for
  depression detection,'' in \emph{ICASSP}.\hskip 1em plus 0.5em minus
  0.4em\relax IEEE, 2019, pp. 6525--6529.

\bibitem{yang2012detecting}
Y.~Yang \emph{et~al.}, ``Detecting depression severity from vocal prosody,''
  \emph{IEEE transactions on affective computing}, vol.~4, no.~2, pp. 142--150,
  2012.

\bibitem{afshan2018effectiveness}
A.~Afshan \emph{et~al.}, ``Effectiveness of voice quality features in detecting
  depression,'' \emph{Interspeech}, 2018.

\bibitem{seneviratne2022multimodal}
N.~Seneviratne and C.~Espy-Wilson, ``Multimodal depression classification using
  articulatory coordination features and hierarchical attention based text
  embeddings,'' in \emph{ICASSP}.\hskip 1em plus 0.5em minus 0.4em\relax IEEE,
  2022, pp. 6252--6256.

\bibitem{yang2020feature}
L.~Yang \emph{et~al.}, ``Feature augmenting networks for improving depression
  severity estimation from speech signals,'' \emph{IEEE Access}, vol.~8, pp.
  24\,033--24\,045, 2020.

\bibitem{vazquez2020automatic}
A.~V{\'a}zquez-Romero \emph{et~al.}, ``Automatic detection of depression in
  speech using ensemble convolutional neural networks,'' \emph{Entropy},
  vol.~22, no.~6, p. 688, 2020.

\bibitem{harati2021speech}
A.~Harati \emph{et~al.}, ``Speech-based depression prediction using
  encoder-weight-only transfer learning and a large corpus,'' in
  \emph{ICASSP}.\hskip 1em plus 0.5em minus 0.4em\relax IEEE, 2021, pp.
  7273--7277.

\bibitem{wang22z_interspeech}
J.~Wang \emph{et~al.}, ``{Unsupervised Instance Discriminative Learning for
  Depression Detection from Speech Signals},'' in \emph{Proc. Interspeech},
  2022, pp. 2018--2022.

\bibitem{di2021using}
Y.~Di \emph{et~al.}, ``Using i-vectors from voice features to identify major
  depressive disorder,'' \emph{Journal of Affective Disorders}, vol. 288, pp.
  161--166, 2021.

\bibitem{egas2022automatic}
J.~V. Egas-L{\'o}pez \emph{et~al.}, ``Automatic assessment of the degree of
  clinical depression from speech using x-vectors,'' in \emph{ICASSP}.\hskip
  1em plus 0.5em minus 0.4em\relax IEEE, 2022, pp. 8502--8506.

\bibitem{ravi2022fraug}
V.~Ravi \emph{et~al.}, ``Fraug: A frame rate based data augmentation method for
  depression detection from speech signals,'' in \emph{ICASSP}.\hskip 1em plus
  0.5em minus 0.4em\relax IEEE, 2022, pp. 6267--6271.

\bibitem{dumpala2022detecting}
S.~H. Dumpala \emph{et~al.}, ``Detecting depression with a temporal context of
  speaker embeddings,'' \emph{Proc. AAAI SAS}, 2022.

\bibitem{lustgarten2020digital}
S.~D. Lustgarten \emph{et~al.}, ``Digital privacy in mental healthcare: current
  issues and recommendations for technology use,'' \emph{Current opinion in
  psychology}, vol.~36, pp. 25--31, 2020.

\bibitem{suhasprivacy}
B.~Suhas \emph{et~al.}, ``Privacy sensitive speech analysis using federated
  learning to assess depression,'' \emph{ICASSP}, 2022.

\bibitem{dumpala21_smm}
S.~H. Dumpala \emph{et~al.}, ``{Sine-Wave Speech and Privacy-Preserving
  Depression Detection},'' in \emph{Proc. SMM21, Workshop on Speech, Music and
  Mind 2021}, 2021, pp. 11--15.

\bibitem{li2020speaker}
H.~Li \emph{et~al.}, ``Speaker-invariant affective representation learning via
  adversarial training,'' in \emph{ICASSP}.\hskip 1em plus 0.5em minus
  0.4em\relax IEEE, 2020, pp. 7144--7148.

\bibitem{gat2022speaker}
I.~Gat \emph{et~al.}, ``Speaker normalization for self-supervised speech
  emotion recognition,'' in \emph{ICASSP}.\hskip 1em plus 0.5em minus
  0.4em\relax IEEE, 2022, pp. 7342--7346.

\bibitem{yin2020speaker}
Y.~Yin \emph{et~al.}, ``Speaker-invariant adversarial domain adaptation for
  emotion recognition,'' in \emph{Proceedings of the 2020 International
  Conference on Multimodal Interaction}, 2020, pp. 481--490.

\bibitem{ravi22_interspeech}
V.~Ravi \emph{et~al.}, ``{A Step Towards Preserving Speakers’ Identity While
  Detecting Depression Via Speaker Disentanglement},'' in \emph{Proc.
  Interspeech}, 2022, pp. 3338--3342.

\bibitem{chen2022wavlm}
S.~Chen \emph{et~al.}, ``Wavlm: Large-scale self-supervised pre-training for
  full stack speech processing,'' \emph{IEEE Journal of Selected Topics in
  Signal Processing}, vol.~16, pp. 1505--1518, 2022.

\bibitem{desplanques2020ecapa}
B.~Desplanques \emph{et~al.}, ``Ecapa-tdnn: Emphasized channel attention,
  propagation and aggregation in tdnn based speaker verification,'' in
  \emph{Proc. Interspeech}, 2020, pp. 3830--3834.

\bibitem{valstar2016avec}
M.~Valstar \emph{et~al.}, ``Avec 2016: Depression, mood, and emotion
  recognition workshop and challenge,'' in \emph{Proceedings of the 6th
  international workshop on AVEC}, 2016, pp. 3--10.

\bibitem{ma2016depaudionet}
X.~Ma \emph{et~al.}, ``Depaudionet: An efficient deep model for audio based
  depression classification,'' in \emph{Proceedings of the 6th international
  workshop on audio/visual emotion challenge}, 2016, pp. 35--42.

\bibitem{speechbrain}
M.~Ravanelli \emph{et~al.}, ``{SpeechBrain}: A general-purpose speech
  toolkit,'' 2021, arXiv:2106.04624.

\bibitem{Morais_ser_ecapa}
E.~Morais \emph{et~al.}, ``Speech emotion recognition using self-supervised
  features,'' in \emph{ICASSP}, 2022, pp. 6922--6926.

\bibitem{bailey2021gender}
A.~Bailey \emph{et~al.}, ``Gender bias in depression detection using audio
  features,'' in \emph{2021 29th EUSIPCO}.\hskip 1em plus 0.5em minus
  0.4em\relax IEEE, 2021, pp. 596--600.

\bibitem{wang22q_interspeech}
D.~Wang \emph{et~al.}, ``{ECAPA-TDNN Based Depression Detection from Clinical
  Speech},'' in \emph{Proc. Interspeech}, 2022, pp. 3333--3337.

\bibitem{schilling2021quantifying}
A.~Schilling \emph{et~al.}, ``Quantifying the separability of data classes in
  neural networks,'' \emph{Neural Networks}, vol. 139, pp. 278--293, 2021.

\bibitem{huang2020exploiting}
Z.~Huang \emph{et~al.}, ``Exploiting vocal tract coordination using dilated
  cnns for depression detection in naturalistic environments,'' in
  \emph{ICASSP}.\hskip 1em plus 0.5em minus 0.4em\relax IEEE, 2020, pp.
  6549--6553.

\bibitem{jiang2020speech}
D.~Jiang \emph{et~al.}, ``Speech simclr: Combining contrastive and
  reconstruction objective for self-supervised speech representation
  learning,'' in \emph{Proc. Interspeech}, 2021, pp. 1544--1548.

\bibitem{oord2018representation}
A.~V.~D. Oord \emph{et~al.}, ``Representation learning with contrastive
  predictive coding,'' \emph{Proc. of NIPS}, 2018.

\bibitem{chen22_interspeech}
W.~Chen \emph{et~al.}, ``{SpeechFormer: A Hierarchical Efficient Framework
  Incorporating the Characteristics of Speech},'' in \emph{Proc. Interspeech
  2022}, 2022, pp. 346--350.

\bibitem{feng2022toward}
K.~Feng \emph{et~al.}, ``Toward knowledge-driven speech-based models of
  depression: Leveraging spectrotemporal variations in speech vowels,'' in
  \emph{IEEE-EMBS ICBHI}.\hskip 1em plus 0.5em minus 0.4em\relax IEEE, 2022,
  pp. 01--07.

\bibitem{eyben2010opensmile}
F.~Eyben \emph{et~al.}, ``Opensmile: the munich versatile and fast open-source
  audio feature extractor,'' in \emph{Proc. of the 18th ACM international
  conference on Multimedia}, 2010, pp. 1459--1462.

\end{thebibliography}

\end{document}